\documentclass[%
 aip,
 jmp,%
 amsmath,amssymb,
preprint,%
]{revtex4-1}

\usepackage{graphicx}
\usepackage{dcolumn}
\usepackage{bm}
\usepackage{graphicx}
\usepackage{amsmath, amssymb, bm, mathptmx}
\usepackage{textgreek}
\usepackage{tikz}
\usepackage{pgfplots, pgfplotstable}
\usepackage[T1]{fontenc} 
\usepackage{geometry} 
\usepackage{xcolor}
\usepackage [english]{babel}
\usepackage{array}
\usepackage{enumitem}
\usepackage{tabularx}
\usepackage{multirow}
\usepackage{anyfontsize}
\usepackage{url}

\pgfplotsset{compat=newest} 
\usepgfplotslibrary{units} 
\usetikzlibrary {positioning, backgrounds, calc, arrows}

\begin{document}


\title{Symmetries in the time-averaged dynamics of networks:\\
reducing unnecessary complexity through minimal network models}

\author{Francesco Sorrentino\textsuperscript{1}, Abu Bakar Siddique\textsuperscript{1},and  Louis M. Pecora\textsuperscript{2}\\ \small \textsuperscript{1)}\emph{Department of Mechanical Engineering, The University of New Mexico}
	\\ \small \textsuperscript{2)}\emph{Code 6343, Naval Research Laboratory, Washington, D.C. 20375}}

\date{\today}

\begin{abstract}
Complex networks are the subject of fundamental interest from the scientific community at large. Several metrics have been introduced to characterize the structure of these networks, such as the degree distribution, degree correlation, path length, clustering coefficient, centrality measures etc. Another important feature is the presence of network symmetries. In particular, the effect of these symmetries has been studied in the context of network synchronization, where they have been used to predict the emergence and stability of cluster synchronous states. Here we provide theoretical, numerical, and experimental evidence that network symmetries play a role in a substantially broader class of dynamical models on networks, including epidemics, game theory, communication, and coupled excitable systems. Namely, we see that in all these models, nodes that are related by a symmetry relation show the same time-averaged dynamical properties. This discovery leads us to propose reduction techniques for exact, yet minimal, simulation of complex networks dynamics, which we show are effective in order to optimize the use of computational resources, such as computation time and memory.
\end{abstract}

\maketitle

\begin{quotation}
There has been substantial research into the role of the network symmetries in affecting cluster synchronization. This paper discusses the effects of the network symmetries on other types of network dynamics, including evolutionary games, traffic models, and neuronal dynamical models.
\end{quotation}

	\section{Outline} \label{sec:oultine}
	In recent years a large body of  research has investigated the dynamics of complex networks, including percolation \cite{moore2000epidemics}, epidemics \cite{ganesh2005effect, earn2000simple}, synchronization \cite{SUCNS, pecora1990synchronization,Ro:Pi:Ku,belykh2005synchronization1,belykh2005synchronization2,belykh2011mesoscale,belykh2001cluster,taylor2009dynamical,tinsley2012chimera,nature_fs}, evolutionary games \cite{nature_gt, weibull1997evolutionary}, neuronal models \cite{luke2013complete,uzuntarla2017inverse}, and traffic dynamics \cite{korea,korea3,So:Va,Oh:Sa,Mo:Pa02,Gu:Gu,fs_comm_01, fs_comm_02, fs_comm_03}. These studies are relevant to better model, understand, design, and control networks in technological, biological, and social applications. Extensive research has shown that the topology of these networks (e.g. their degree distribution \cite{pastor2001epidemic,barabasi1999emergence}, degree correlation \cite{newman2003mixing,newman2002assortative}, community structure \cite{girvan2002community}, etc.) plays a significant role in their dynamical time evolution \cite{boccaletti2006complex}. 
		
	\begin{figure}[h!t!]
		\centering
\includegraphics[width=0.9\textwidth]{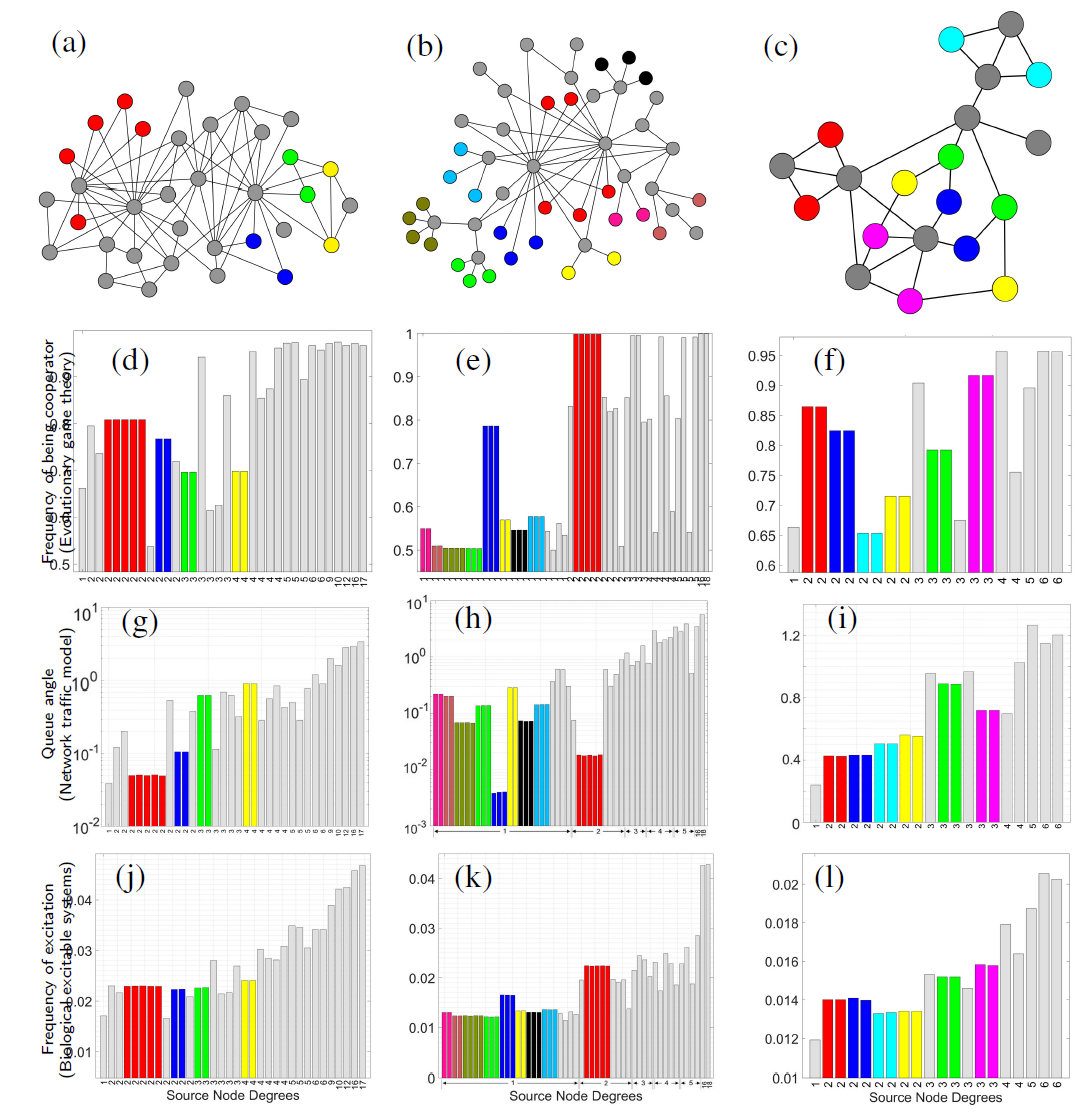}
		\caption{\textbf{Effects of the network symmetries in three different dynamical models/networks:} (a) the Zachary's Karate Club network \cite{zachary1977information}, (b) the Bellsouth network \cite{dataset} and (c) a random graph, nodes colored the same are in the same symmetry cluster except the gray colored nodes, each of which is in a cluster by itself. (d-f) Prisoner's Dilemma game played, (g-i) Network traffic model simulated and (j-l) dynamics of the Kinouchi Copelli model \cite{kc_model} in these networks (a-c).}
		\label{fig:AB}
	\end{figure}
	
	Another important feature of the network topology is the presence of network symmetries, which so far have remained to some extent unexplored, with some exceptions \cite{nature_fs, nicosia2013remote, golubitsky2003symmetry, golubitsky2012singularities, NCs}. These symmetries have been shown to be commonplace in many real networks \cite{macarthur2009spectral}, hence it becomes important to understand how they can affect the dynamics of the network. References \cite{nature_fs, nicosia2013remote, golubitsky2003symmetry, golubitsky2012singularities, NCs} have focused on the role played by the network symmetries on the emergence of cluster synchronization. Here we consider several well known dynamical models on networks and try to illustrate the effects of the underlying network symmetries on the network dynamics. Our study indicates that network symmetries play a role in all the dynamical models considered and in particular: evolutionary games  \cite{nature_gt, weibull1997evolutionary}, network traffic  \cite{korea,korea3,So:Va,Oh:Sa,Mo:Pa02,Gu:Gu,fs_comm_01, fs_comm_02, fs_comm_03}, and propagation of excitation among excitable systems \cite{KI1,KI2,KI3};  and thus suggests the effects of the symmetries on the dynamics may be a rather general feature of complex networks. However, as we will see, the particular effect of the symmetries varies based on the particular type of dynamics considered. 
	
	
	Here, the topology of a network is described by the adjacency matrix $A=\{A_{ij}\}$, where $A_{ij}=A_{ji}$ is equal to $1$ if node $j$ and $i$ affect each other and is equal to $0$ otherwise. We define a network symmetry as a permutation of the network nodes that leaves the network structure unaltered. The symmetries of the network form a (mathematical) group $\mathcal{G}$. Each element of the group can be described by a permutation matrix $\Pi$ that re-orders the nodes in a way that leaves the network structure unchanged (that is, each $\Pi$ commutes with $A$, $\Pi A = A\Pi$). Though the set of symmetries (or automorphisms) of a network can be quite large, even for small networks, it can be computed from knowledge of the matrix $A$ by using widely available discrete algebra routines. In fact, except for simple cases for which it may be possible to identify the symmetries by inspection, in general for an arbitrary network, the use of a software is required. In this study we used Sage \cite{sage}, an open-source mathematical software. Once the symmetries are identified, the nodes of the network can be partitioned into $M$ \textit{clusters} by finding the orbits of the symmetry group, i.e., the disjoint sets of nodes that, when all of the symmetry operations are applied, permute among one another in the same set.

    \section{Symmetries of the Network Dynamics} \label{sec:snd}
    Figure\ \ref{fig:AB}(a), (b), and (c) shows three examples of undirected networks: the Zachary's Karate Club network \cite{zachary1977information} of $N=34$ nodes, the Bell South network \cite{knight2011internet} of $N=51$ nodes and a randomly generated Erd\"os-R\'enyi (ER) graph of $N=20$ nodes, respectively. Each node of the Karate Club network is a member of a university karate club and a connection represents a friendship relation between the members. The nodes of the Bell South network are the IP/MPLSs (Multiprotocol Label Switching: a switching mechanism used in high-performance telecommunications networks) backbone and connections represent routing paths. In Fig.\ \ref{fig:AB}(a), (b) and (c) colors of the nodes indicate the clusters they belong to, either non-trivial (i.e. clusters with more than one node in them) or trivial clusters (clusters with only one node in them). All the nodes in trivial clusters are colored gray, while the non-trivial clusters are colored differently. The Karate Club network in Fig.\ \ref{fig:AB}(a) has $C=4$ nontrivial clusters, $23$ trivial clusters and $480$ symmetries. The Bell South network in Fig.\ \ref{fig:AB}(b) displays $C=9$ nontrivial clusters, $24$ trivial clusters and $29,859,840$ symmetries. The random network in Fig.\ \ref{fig:AB}(c) has $C=6$ non-trivial clusters, $8$ trivial clusters and $8$ symmetries. The rest of Fig.\ \ref{fig:AB} has a total of nine panels, one for each of three networks and each of three dynamical models. The three dynamical models are: evolutionary game theory (d-f), network traffic (g-i), and propagation of excitation among excitable system (j-l). 

We now briefly introduce the models, which are all stochastic in nature. The evolutionary game theory dynamics models the evolution of cooperation and defection in a population of coupled agents (nodes), playing the Prisoner's Dilemma game. At each time step a node is randomly selected and its strategy is updated. The new strategy to be adopted is probabilistically determined based on the payoffs of the nodes surrounding the selected node and their strategy selection. 

	Each one of the network nodes (agents) iteratively plays a version of the Prisoner's Dilemma game \cite{nature_gt}. Each node $i$ can either be a cooperator ($S_{i}=1$) or a defector ($S_{i}=0$). The network connectivity is defined by the adjacency matrix $A$, described earlier. We define a payoff between two players, based on the well known \emph{Prisoner's Dilemma} game. There are two types of strategy adopted by the players: \emph{cooperation} and \emph{defection}. A cooperator pays a cost $c$ for each one of the agents it is connected to and a defector pays nothing \cite{nature_gt}. Each node receives a benefit equal to $b$ for each cooperator it is connected to. When playing the game, node $i$ receives a payoff equal to 
    
    \begin{equation}
    \xi_{i} = \sum_{j}(A_{ij}bS_{j}- A_{ji}cS_{i}).
    \end{equation}
    
    We define the fitness \cite{nature_gt} of each node to be $f_{i}=1-\omega+\omega \xi_{i}$, where $0 \leq \omega \leq 1$ measures the intensity of selection:  $\omega\simeq 1$ means strong selection, that is the fitness is almost equal to the payoff and $\omega\simeq 0$ means weak selection, that is the fitness is almost independent of the payoff and close to $1$. The literature \cite{nature_gt, wu2007cooperation, wang2010evolutionary, tan2014structure} focuses on the case of weak selection, which is also what we consider here (in all our simulations we either set $\omega=0.1$ or $\omega=0.2$). Following \cite{nature_gt} we choose a `Death-birth' (DB) updating rule for the game evolution. Namely, in each time step a randomly selected node $i$ is replaced by a new offspring (node). The new offspring evolves into either a cooperator or a defector depending on the fitness of the surrounding agents. We set the probability of that new node to be a cooperator to be $\sigma(F_{Ci}-F_{Di})$, where $F_{Ci}$ and $F_{Di}$ are the fitnesses of cooperators and defectors in the neighboring nodes and $\sigma$ is a monotonically increasing function such that $0\leq\sigma\leq 1$. This reflects a higher propensity of turning into a cooperator based on how \emph{well} the neighbors of a given node that are cooperators are doing with respect to the other neighbors of that node that are defectors. The total fitness of the neighbors of player $i$ is equal to
 	\begin{equation}
 	F_{i}=\sum\limits_{j}^{N}A_{ij}f_{j}
 	\label{eq:total_fitness}
 	\end{equation} 
 	The total fitness of the cooperators, $F_{Ci}$ and defectors, $F_{Di}$ in the neighboring nodes of $i$ is defined as,
 	\begin{equation}
 	\begin{split}
 	F_{Ci} =&\sum\limits_{j}^{N}A_{ij}S_{j}f_{j} = \sum_{j}A_{ij}S_{j}(1-\omega)+\omega\sum_{j}A_{ij}S_{j}\xi_{j} \cr
 	F_{Di} =& F_i - F_{Ci} = \sum_{j}^{N}A_{ij}(1-S_{j})(1-\omega)+\omega\sum_{j}A_{ij}(1-S_{j})\xi_{j}
 	\end{split}
 	\label{eq:fitness_cooperator}
 	\end{equation}

 	Letting $x_i = (F_{Ci}-F_{Di})$, we write the probability that the new offspring will be a cooperator $\sigma(x_i)$. Here we set $\sigma(x_i) = \gamma x_i + \epsilon$, where $\gamma>0$ and $\epsilon$ are two arbitrary constants. In all our numerical simulations the values of $\gamma$ and $\epsilon$ were chosen so as to ensure $0\leq \sigma \leq 1$ for all $i$'s.  In Sec.\ S1 of the Supplementary Information we obtain a set of equations that describe the time evolution of the  game  and prove its  equivariance with respect to permutations of the network nodes that are  in the automorphism group of $A$.
    
	We numerically iterated the game on several networks, including the three shown in Figs.\ \ref{fig:AB} (a), (b) and (c) for a number of time-steps and for each node $i$ we monitored $\left<S_i\right>$ the fraction of times a node spends in the cooperator state. For each run, the game was iterated until a state was reached in which the number of cooperators and defectors stabilized. Figures \ref{fig:AB}(d), (e) and (f) show $\left<S\right>_i$, the fraction of times that each node spends in the cooperator state for each one of the nodes of the networks in Fig.\ \ref{fig:AB}(a), (b) and (c), respectively.


In the network traffic model, packets are originated at source nodes, get routed through a sequence of intermediate nodes, until they reach the destination nodes and get removed from the network \cite{korea,korea3,So:Va,Oh:Sa,Mo:Pa02,Gu:Gu,fs_comm_01, fs_comm_02, fs_comm_03}. At every intermediate node, packets are placed at the bottom of that node's queue.  When they reach the top of the queue they get routed to one of the neighboring nodes. Here we consider a simple routing strategy that attempts to avoid nodes with large queues assigning them a lower probability of being selected for routing. Figure 1(g), (h), and (i) show the rate of growth of the queue length (number of packets in the queue) at each node for the three networks shown in Figure 1(a), (b), and (c), respectively. 

Finally, we consider a network of coupled excitable systems \cite{KI1}. Each one of these systems  can be in either one of three states: quiescent, excited, and refractory. Nodes that are excited can excite neighboring nodes that are in the quiescent state with a certain probability. Figure 1(j), (k), and (l) show the frequency of excitation at each node for the three networks shown in Figure 1(a), (b), and (c), respectively.   A more precise and  detailed description of the evolutionary game theory model is provided in Supplementary Information Sec.\ S1, of the network traffic model in Supplementary Information Sec.\ S2, and of the excitable systems model in Supplementary Information Sec.\ S3.

Our main result is that for all the three networks and the three dynamical models, \emph{nodes that belong to the same cluster show the same time-averaged dynamics.} This is illustrated in detail in Fig.\ 1 (d)-(l). While this observation holds irrespective of the particular network and type of dynamics, the particular time-averaged value attained by the nodes in the same cluster is network and model specific. Note that in the figure nodes are ordered by their degree (which is the label on the abscissa-axis). For example, for the game theory model, we observe that the nodes in the same cluster approximately show the same frequency of being a cooperator (or defector) but that does not necessarily correlate with the degree. Here we see that the symmetries in the network topology can predict dynamics better than the nodes' degrees. 

For each one of the dynamical models considered, we have performed an analysis to: (i) predict the emergence of clusters when the dynamics is averaged in time and (ii)  predict the values attained by the nodes in each cluster. The results of this analysis is reported for the evolutionary game, communication model and the excitable systems model in the Supplementary Information, Secs.\ S1, S2, and S3.

	
	
	\section{Quotient Graph Reduction} \label{sec:qtgr}
	As mentioned in the introduction, symmetries are common features of biological networks, technological networks, social networks etc. MacArthur \textit{et al}.\ \cite{macarthur2008symmetry} have analyzed datasets of large complex networks and have found that these present large numbers of symmetries, see  Supplementary Information Sec.\ S4. Intensive research in social sciences, biology, engineering, and physics uses numerical simulations of large complex networks to understand and predict their dynamical behavior (e.g., in a given network, the critical value of the infection rate above which an epidemics occurs), in order to better characterize and control real-world phenomena.

 	Our results in Secs.\ \ref{sec:snd}, S1, S2, and S3 point out that nodes that are related by a symmetry operation display the same time-averaged dynamical behavior. This immediately raises the question whether a reduction of the dynamics is possible in which \emph{duplicate nodes} can be omitted, leading to \emph{minimal models} of complex networks, and so to a better exploitation  of computational resources in numerical simulations, such as computation time and memory.  Related questions have been asked in the  literature of complex networks, where nodes have been grouped according to some of their features, most notably the degree \cite{pastor2001epidemic} and these approaches have been successful at predicting and explaining several network properties, in particular in the case of scale free networks \cite{barabasi1999emergence}. While these approaches are typically based on mean-field models and thus approximate, here our grouping of nodes is based on the exact concept of a symmetry. 
 	
 Our ultimate goal is to generate {minimal network models} that reproduce certain features of the dynamics by using the least possible number of nodes. In the case of synchronization dynamics, we know that a \emph{quotient network reduction} is possible, in which the exact cluster-synchronous time-evolution is generated by a minimal number of nodes (i.e., one node for each cluster).  However, how to obtain minimal network models for other types of dynamics, remains an open question. To address this issue, here we will briefly review the concept of a \emph{quotient network}.
 
Under the action of the symmetry group, the set of the network nodes is partitioned  into $C$ disjoint structural equivalence classes called the group  \emph{orbits},   $\mathcal{O}_1,\mathcal{O}_2,...,\mathcal{O}_C$, such that $\bigcup\limits_{\ell =1}^{C}|\mathcal{O}_\mathcal{G}^{\ell}| = N \text{ and } \mathcal{O}_\mathcal{G}^i \cap \mathcal{O}_\mathcal{G}^j=0, \text{ where } i,j=1,2,...,C, j\neq i$.

 Then we can define a $C \times C$ matrix $\hat{A}$ corresponding to the quotient network such that for each pair of sets $(\mathcal{O}^v, \mathcal{O}^u)$,
 	\begin{equation}
	\hat{A}_{uv}= \sum_{j \in O^v} A_{ij},
 	\end{equation}
 	for any $i \in O^u$ (i.e., independent of $i \in O^u$), and for $u,v = 1,2,...,C$.


\begin{figure}[ht!]
	\centering
		 		\includegraphics[width=0.8\textwidth]{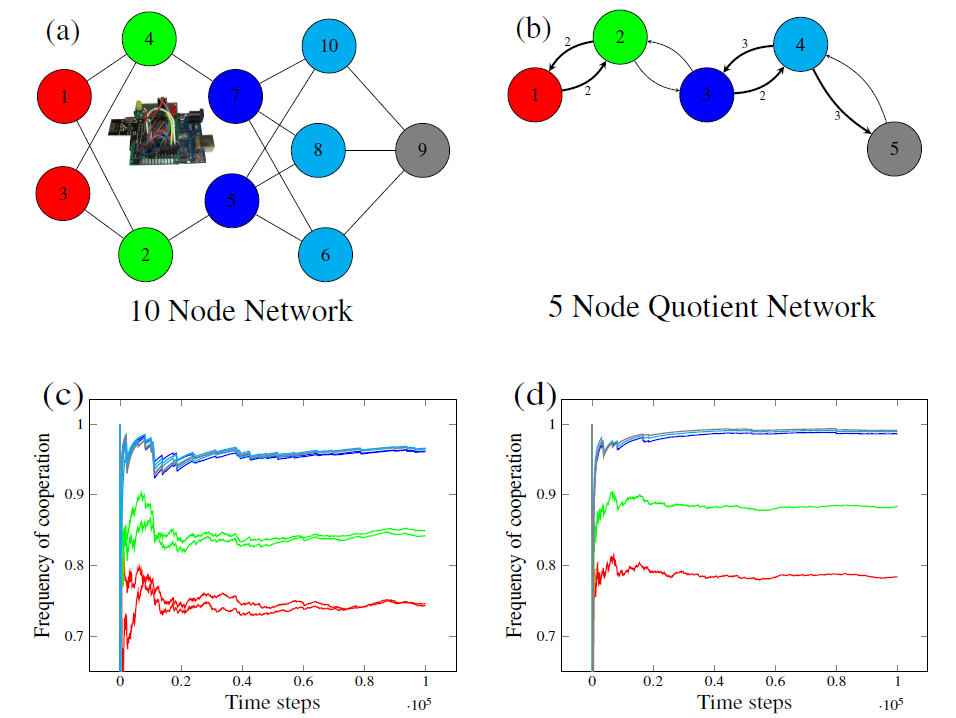}
	\caption{(a) A 10 node network and (b) its 5 node quotient reduction. Inset of (a) shows one arduino board with radio transmitter, which we used in our experiments. Plots (c) and (d) are experimental time traces showing the running average for the fraction of times each node spends in the cooperator state for the full network in (a) and the quotient network in (b), respectively. Colors in (c) and (d) are consistent with (a) and (b).}
	\label{fig:experiment}
\end{figure}

	In Fig. \ref{fig:experiment}(a) we show a randomly generated network of 10 nodes and in Fig. \ref{fig:experiment}(b) its quotient graph reduction. We see that all the nodes in the same cluster (these are colored the same in the figure on the left), map to only one node of the quotient network (on the right) and the color identifies the reduced node. Note that the quotient network may be directed even if the original full network is undirected. It is also possible that nodes of the quotient network form self connections, which represent connections between nodes belonging to the same cluster in the original graph.
 	
 	In general, the mathematical equations we have derived in Sec.\ \ref{sec:snd} and S1, S2, S3 of Supplementary Information for all the three models can be projected onto the corresponding quotient network equations (see the Supplementary Information Sec.\ S4
 	for an example). However, obtaining an equivalent model that can be simulated on the quotient network and reproduce the original full network dynamics may require a particular adaptation of the model, which is model specific.
 	
 	We show that the quotient graph can be conveniently exploited in simulations involving large networks to reduce computation time and memory. These simulations may be used to model various type of dynamics, including epidemics, congestion, emergence of cooperation, as discussed in Sec.\ \ref{sec:snd}, just to mention a few examples.  In order to demonstrate this point we consider the evolutionary game theory model presented in Sec.\ \ref{sec:snd} and for a number of networks, we study how well the corresponding quotient graphs can approximate the  evolutionary dynamics of the original full network. In what follows we describe evolution of the Prisoner's Dilemma dynamics, as described in Sec.\ \ref{sec:snd}, on the quotient network. We indicate with $\hat{S}_j = \{0,1\}$ the strategy of node $j$ of the quotient network, where $0$ represents defection and $1$ represents cooperation. At each iteration of the game, the payoff received by quotient node $i$ from its neighboring nodes is equal to $\hat{\xi}_i  = \sum_j \hat{A}_{ij} \left(b\hat{S}_j-c\hat{S}_i\right)$. Note that this expression for the payoff differs from that for the full network in Eq.\ (1) as the mapping of nodes from the full network to the quotient network preserves the indegree of the nodes, but not the outdegree. 
	Moreover, the fitness of quotient node $i$ is equal to $\hat{f}_i=1-\omega+\omega\hat{\xi}_i$. Similarly we write the expressions, 
	\begin{equation}
	\begin{split}
	\hat{F}_i &=\sum_{j}\hat{A}_{ij}\hat{f}_j \cr
	\hat{F}_{Ci} &=\sum_{j}\hat{A}_{ij}\hat{S}_{j}(1-\omega)+\omega\sum_{j}\hat{A}_{ij}\hat{S}_{j}\hat{\xi}_{j} \cr
	\hat{F}_{Di} &=\sum_{j}\hat{A}_{ij}(1-\hat{S}_{j})(1-\omega)+\omega\sum_{j}\hat{A}_{ij}(1-\hat{S}_{j})\hat{\xi}_{j}
	\end{split}
	\end{equation}
	where $\hat{F}_i$, $\hat{F}_{Ci}$ and $\hat{F}_{Di}$ are the total fitness, the fitness of the neighboring cooperators, and the fitness of the neighboring defectors of a quotient node $i$, respectively.
	We set the same cost $c$ and benefit $b$ as for the full network. At each time step, a node of the quotient network is selected with probability proportional to the cardinality of its orbit set and replaced by a new offspring. This new node becomes a cooperator with a probability $\sigma\left(\hat{F}_{Ci} - \hat{F}_{Di}\right)$, where $\sigma$ is the  function defined in Sec.\ 2. 
 	
\begin{figure}[ht!]
	\centering
	\includegraphics[width=\textwidth]{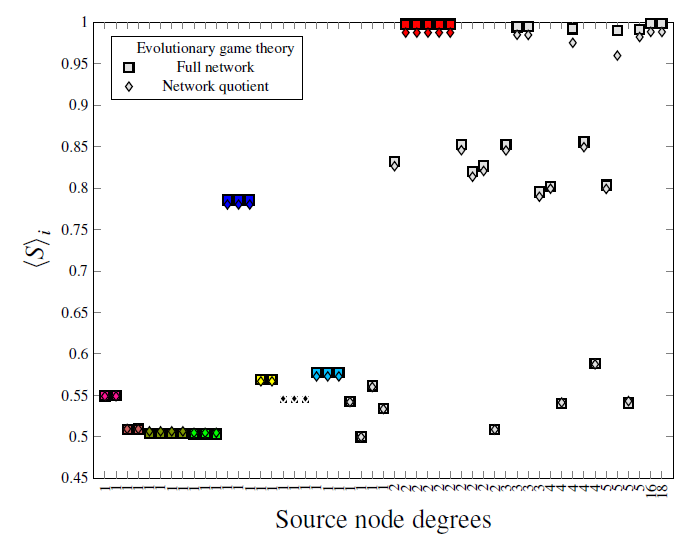} \label{fig:sim_match_quotient} 
\caption{Comparison between simulation results of the evolutionary game on the Bellsouth network
in Fig.\ 1(b) and its quotient version. The diamond marks correspond to simulation on the full
network and the square marks correspond to the simulation on the quotient network. Points that
are colored the same correspond to nodes in the same cluster (coloring is consistent with Fig.\ 1(a)
and (b), except for the gray colored nodes, each of which is in a cluster by itself).}
\end{figure}

 	We expect that the time averages $\left<\xi_f\right>\approx\left<\hat{\xi}_q\right>$, $\left<F_{Cf}\right>\approx\left<\hat{F}_{Cq}\right>$ and  $\left<F_{Df}\right>\approx\left<\hat{F}_{Dq}\right>$, where node $f$ of the full network maps to node $q$ of the quotient network (for nodes of the full network that are in the same cluster, we know that the time averages are the same from the analysis in Sec.\ \ref{sec:snd}).
 	Moreover, the time-averaged strategy for a node of the quotient network $\left<\hat{S}_q\right>$ is the same as the time-averaged strategy for the corresponding node of the full network, $\left<S_f\right>$. Figure \ref{fig:sim_match_quotient} compares the simulation results  for the Bellsouth network (squares) and its quotient reduction (diamonds).  As can be seen, we find very good agreement between the full and quotient network time-averaged dynamics. This indicates that, if we are interested in the time-averaged behavior, we can equivalently perform a simulation on the full network or on the reduced quotient network.

In order to test the robustness of our results in a real setting, we have built an experimental network of coupled agents iteratively playing the Prisoner's Dilemma. This network is composed of 10 wirelessly coupled transceiver modules (details in Supplementary Sec.\ S7) as shown in Fig.\ \ref{fig:experiment}a using nRF24L01 2.4 GHz RF transceivers on Arduino boards. The transceiver modules act as the nodes of the network. All the transceiver modules have unique addresses, i.e., communication is one to one. To construct this network, communication links
of each module are restricted as per the topology of the network, i.e., only the connected modules can communicate
and share information with each other. For example, radio module 7 in Fig. \ref{fig:experiment}a  can only communicate with modules 4, 6, 8, and 10. 
 This experimental realization is subject to practical limitations that are hard to reproduce in simulation. In particular, in our experimental setting these limitations are mainly imposed by the reliability of the radio communication between the individual units (details in the Supplementary Information Sec.\ S7).   We have also built an experimental version of the quotient network in Fig.\ \ref{fig:experiment}b. 
The experimental time traces for the networks in Fig.\  \ref{fig:experiment} a and b when the game is played are shown in Fig.\ \ref{fig:experiment}c and d, respectively. We see that: i) nodes of the full network that are in the same cluster  attain the same  frequency of cooperation as the game is iterated and ii)  for large time the quotient network well predicts the full network experimental dynamics.

	Since the quotient graphs have in general fewer nodes than the full graphs, they can be advantageous in terms of both memory and time needed in simulation. The size of the adjacency matrix reduces from $N \times N$ to $C \times C$ and the number of nonzero elements of the matrix decreases from $NK_{av}$ to roughly $\sqrt{NC}K_{av}$, where $K_{av}$ is the average node degree of the full network. We define the reduction coefficient $\rho_g=C/N$. A smaller value of the ratio $\rho_g$ indicates higher reduction of the number of nodes in the quotient graph with respect to the original graph.
	Note that a critical aspect of simulation of large real networks is the limitation of software memory allocation. We computed the CPU time required by a single iteration of the prisoner's dilemma dynamics for several networks and their quotients.  Fig.\ \ref{fig:error_result}(a) and Table S2 in the Supplementary Information show the CPU time ratio $\rho_t = t_q/t_f$, where $t_f$ and $t_q$ are the CPU time for an iteration of full and quotient network, respectively.

	For both the full network and the quotient network we also computed the simulation convergence time, which we measured as follows. We randomly picked an initial condition for each node of the quotient network and assigned the same initial condition to all the nodes in the corresponding  cluster of the full network. At each time step, we computed the running average for the fraction of times a node spent in the cooperation state. For each node we measured its individual convergence time, i.e., the time after which the running average remained steadily in a $[- \delta, +\delta]$ interval of the final state. The convergence time of the network was taken to be the largest of the convergence times of the nodes.     Fig.\ \ref{fig:error_result}(b) and  Table S2 in the Supplementary Information show the convergence time ratio $\rho_c = \tau_q/\tau_f$, where $\tau_f$ and $\tau_q$ are the convergence times for the full and the quotient network, respectively.

\begin{figure}[ht!]
	\centering
	\includegraphics[width=\textwidth]{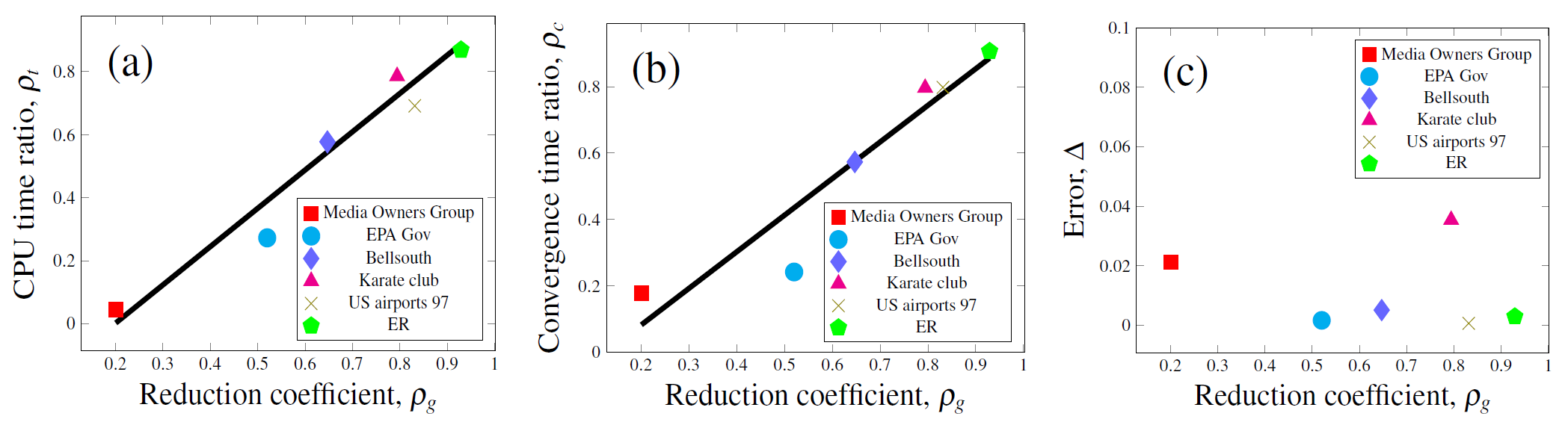}
\caption{(a) CPU time ratio (b) Convergence time ratio (c) Error between the full and quotient network time-averaged dynamics.}
\label{fig:error_result}
\end{figure}

We note that simulating the dynamics on the quotient graph, rather than on the original network, can reduce the computational effort, while producing approximately the same time-averaged dynamics. In Fig.\ \ref{fig:error_result}(c) and  Table S2 in the Supplementary Information we show the accuracy parameter ${\Delta}$, defined as the normalized average difference of the time-averaged frequency of cooperation between the full and the quotient networks, 
\begin{equation}
\Delta = \frac{1}{N}\sum_{i}^{N}\frac{|\langle S \rangle_{i}-\hat{\langle S \rangle}_{i}|}{S_{i}},
\end{equation}
where $\langle S \rangle_{i}$ and $\hat{\langle S \rangle}_{i}$ are the time-averaged frequency of cooperation of node $i$ and its corresponding node in the quotient network, respectively. 

Figure \ref{fig:error_result} shows that as the reduction coefficient is lowered, the CPU time ratio $\rho_t$ and the convergence time ratio $\rho_c$ decrease in a linear fashion but the normalized error parameter $\Delta$ is roughly independent of $\rho_g$. It is important to look at the $y$-axes of the plots in Figs.\ \ref{fig:error_result}(a), (b) and (c).  For example, for a strong reduction coefficient $\rho_g \simeq 0.2$, corresponding to the Media Owners Group network, the CPU time is lowered by roughly $95 \%$ but the normalized error $\Delta$ only increases by approximately $2\%$.


\section{Conclusions} 
By applying a symmetry analysis to a network, we have uncovered clusters of nodes that are structurally and functionally equivalent. This becomes apparent when monitoring the time-averaged state of the nodes in  a variety of network  models (previous work had only focused on the particular case of synchronization dynamics) and  is confirmed in an experimental realization of an evolutionary game played on a network, in the presence of  noise and communication losses. Thus  it appears that the emergence of  symmetry clusters in the time-averaged dynamics of networks is a quite general feature. For the case of the evolutionary game, we obtain a reduction technique for exact, yet minimal, simulation of complex networks dynamics, which produces similar dynamical results, while computation requires less time and memory. However, a generalization of this quotient network reduction to other types of dynamics is nontrivial. The reason is that each dynamical model involves a different set of \emph{rules}, which may be difficult to convert into equivalent rules for the quotient network.  We hope our work will stimulate further research into
reduction techniques that can be applied in a variety of dynamical models. 

\textbf{Supplementary Material}
Supplementary Material for this paper includes 7 sections. The evolutionary game theory model,  network traffic model, and biological excitable system model are described in Sections S1, S2, and S3, respectively. The equations describing the quotient network dynamics are presented in Section S4. Tables for the symmetries of real networks and the computational aspects of the quotient network simulations are included in Sections S5 and S6, respectively. The experimental realization is described in Section S7.

\textbf{Acknowledgement.} 
This work was supported by  the Office of Naval Research through ONR Award No.  N00014-16-1-2637, the National
Science Foundation through NSF grant CMMI- 1727948, NSF grant CRISP- 1541148,
and the Defense Threat Reduction Agency's Basic Research Program under grant No. HDTRA1-13-1-0020. The authors acknowledge help in running the experiment from Fabio Della Rossa, Robert Morris, Jonathan Ungaro, John Padilla, and Shakeeb Ahmad, all from the University of New Mexico.

\bibliographystyle{unsrt}

\begin{thebibliography}{10}

\bibitem{moore2000epidemics}
Cristopher Moore and Mark~EJ Newman.
\newblock Epidemics and percolation in small-world networks.
\newblock {\em Physical Review E}, 61(5):5678, 2000.

\bibitem{ganesh2005effect}
Ayalvadi Ganesh, Laurent Massouli{\'e}, and Don Towsley.
\newblock The effect of network topology on the spread of epidemics.
\newblock In {\em INFOCOM 2005. 24th Annual Joint Conference of the IEEE
  Computer and Communications Societies. Proceedings IEEE}, volume~2, pages
  1455--1466. IEEE, 2005.

\bibitem{earn2000simple}
David~JD Earn, Pejman Rohani, Benjamin~M Bolker, and Bryan~T Grenfell.
\newblock A simple model for complex dynamical transitions in epidemics.
\newblock {\em Science}, 287(5453):667--670, 2000.

\bibitem{SUCNS}
Arkady Pikovsky, Michael Rosenblum, and Jurgen Kurths.
\newblock {\em Synchronization: A Universal Concept in Nonlinear Sciences}.
\newblock Cambridge University Press, 2003.

\bibitem{pecora1990synchronization}
Louis~M Pecora and Thomas~L Carroll.
\newblock Synchronization in chaotic systems.
\newblock {\em Physical review letters}, 64(8):821, 1990.

\bibitem{Ro:Pi:Ku}
M.~G. Rosenblum, A.~S. Pikovsky, and J.~Kurths.
\newblock Phase synchronization of chaotic oscillators.
\newblock {\em Phys. Rev. Lett.}, 76:1804, 1996.

\bibitem{belykh2005synchronization1}
Igor Belykh, Martin Hasler, Menno Lauret, and Henk Nijmeijer.
\newblock Synchronization and graph topology.
\newblock {\em International Journal of Bifurcation and Chaos},
  15(11):3423--3433, 2005.

\bibitem{belykh2005synchronization2}
Igor Belykh, Enno de~Lange, and Martin Hasler.
\newblock Synchronization of bursting neurons: What matters in the network
  topology.
\newblock {\em Physical review letters}, 94(18):188101, 2005.

\bibitem{belykh2011mesoscale}
Igor Belykh and Martin Hasler.
\newblock Mesoscale and clusters of synchrony in networks of bursting neurons.
\newblock {\em Chaos: An Interdisciplinary Journal of Nonlinear Science},
  21(1):016106, 2011.

\bibitem{belykh2001cluster}
Vladimir~N Belykh, Igor~V Belykh, and Erik Mosekilde.
\newblock Cluster synchronization modes in an ensemble of coupled chaotic
  oscillators.
\newblock {\em Physical Review E}, 63(3):036216, 2001.

\bibitem{taylor2009dynamical}
Annette~F Taylor, Mark~R Tinsley, Fang Wang, Zhaoyang Huang, and Kenneth
  Showalter.
\newblock Dynamical quorum sensing and synchronization in large populations of
  chemical oscillators.
\newblock {\em Science}, 323(5914):614--617, 2009.

\bibitem{tinsley2012chimera}
Mark~R Tinsley, Simbarashe Nkomo, and Kenneth Showalter.
\newblock Chimera and phase-cluster states in populations of coupled chemical
  oscillators.
\newblock {\em Nature Physics}, 8(9):662--665, 2012.

\bibitem{nature_fs}
Louis~M Pecora, Francesco Sorrentino, Aaron~M Hagerstrom, Thomas~E Murphy, and
  Rajarshi Roy.
\newblock Cluster synchronization and isolated desynchronization in complex
  networks with symmetries.
\newblock {\em Nature communications}, 5:304--305, 2014.

\bibitem{nature_gt}
Hisashi Ohtsuki, Christoph Hauert, Erez Lieberman, and Martin~A Nowak.
\newblock A simple rule for the evolution of cooperation on graphs and social
  networks.
\newblock {\em Nature}, 441(7092):502--505, 2006.

\bibitem{weibull1997evolutionary}
J{\"o}rgen~W Weibull.
\newblock {\em Evolutionary game theory}.
\newblock MIT press, 1997.

\bibitem{luke2013complete}
Tanushree~B Luke, Ernest Barreto, and Paul So.
\newblock Complete classification of the macroscopic behavior of a
  heterogeneous network of theta neurons.
\newblock {\em Neural computation}, 25(12):3207--3234, 2013.

\bibitem{uzuntarla2017inverse}
Muhammet Uzuntarla, Ernest Barreto, and Joaquin~J Torres.
\newblock Inverse stochastic resonance in networks of spiking neurons.
\newblock {\em PLoS computational biology}, 13(7):e1005646, 2017.

\bibitem{korea}
K.-I. Goh, B.~Kahng, and D.~Kim.
\newblock Universal behavior of load distribution in scale-free networks.
\newblock {\em Phys. Rev. Lett.}, 87:278701, 2001.

\bibitem{korea3}
K.-I. Goh, B.~Kahng, and D.~Kim.
\newblock Packet transport and load distribution in scale-free network models.
\newblock {\em Physica A}, 318:72--79, 2003.

\bibitem{So:Va}
R.V. Sole and S.Valverde.
\newblock Information transfer and phase transition in a model of data traffic.
\newblock {\em Physica A}, 289(595), 2001.

\bibitem{Oh:Sa}
Toru Ohira and Ryusuke Sawatari.
\newblock Phase transition in a computer network traffic model.
\newblock {\em Physical Review E}, 58(1):193, 1998.

\bibitem{Mo:Pa02}
Y.Moreno, R.Pastor-Satorras, A.Vasquez, and A.Vespignani.
\newblock Critical load and congestion instabilities in scale-free networks.
\newblock {\em cond-mat}, 1(0209474), 2002.

\bibitem{Gu:Gu}
R.~Guimer\`{a}, A.~D\'{i}az-Guilera, F.~Vega-Redondo, A.~Cabrales, and
  A.~Arenas.
\newblock Optimal network topologies for local search with congestion.
\newblock {\em Phys. Rev. Lett.}, 89(24), 2002.

\bibitem{fs_comm_01}
David Arrowsmith, Mario di~Bernardo, and Francesco Sorrentino.
\newblock Communication models with distributed transmission rates and buffer
  sizes.
\newblock In {\em Circuits and Systems, 2006. ISCAS 2006. Proceedings. 2006
  IEEE International Symposium on}, pages 4--pp. IEEE, 2006.

\bibitem{fs_comm_02}
David Arrowsmith, Mario Di~Bernardo, and Francesco Sorrentino.
\newblock Effects of variations of load distribution on network performance.
\newblock In {\em Circuits and Systems, 2005. ISCAS 2005. IEEE International
  Symposium on}, pages 3773--3776. IEEE, 2005.

\bibitem{fs_comm_03}
Mario di~Bernardo and Francesco Sorrentino.
\newblock Network structural properties, communication models and traffic
  dynamics.
\newblock {\em Nolta}, 2006.

\bibitem{pastor2001epidemic}
Romualdo Pastor-Satorras and Alessandro Vespignani.
\newblock Epidemic spreading in scale-free networks.
\newblock {\em Physical review letters}, 86(14):3200, 2001.

\bibitem{barabasi1999emergence}
Albert-L{\'a}szl{\'o} Barab{\'a}si and R{\'e}ka Albert.
\newblock Emergence of scaling in random networks.
\newblock {\em science}, 286(5439):509--512, 1999.

\bibitem{newman2003mixing}
Mark~EJ Newman.
\newblock Mixing patterns in networks.
\newblock {\em Physical Review E}, 67(2):026126, 2003.

\bibitem{newman2002assortative}
Mark~EJ Newman.
\newblock Assortative mixing in networks.
\newblock {\em Physical review letters}, 89(20):208701, 2002.

\bibitem{girvan2002community}
Michelle Girvan and Mark~EJ Newman.
\newblock Community structure in social and biological networks.
\newblock {\em Proceedings of the national academy of sciences},
  99(12):7821--7826, 2002.

\bibitem{boccaletti2006complex}
Stefano Boccaletti, Vito Latora, Yamir Moreno, Martin Chavez, and D-U Hwang.
\newblock Complex networks: Structure and dynamics.
\newblock {\em Physics reports}, 424(4):175--308, 2006.

\bibitem{zachary1977information}
Wayne~W Zachary.
\newblock An information flow model for conflict and fission in small groups.
\newblock {\em Journal of anthropological research}, pages 452--473, 1977.

\bibitem{dataset}
The internet topology zoo.
\newblock \url{http://www.topology-zoo.org/dataset.html}.

\bibitem{kc_model}
Osame Kinouchi and Mauro Copelli.
\newblock Optimal dynamical range of excitable networks at criticality.
\newblock {\em Nature physics}, 2(5):348--351, 2006.

\bibitem{nicosia2013remote}
Vincenzo Nicosia, Miguel Valencia, Mario Chavez, Albert D{\'\i}az-Guilera, and
  Vito Latora.
\newblock Remote synchronization reveals network symmetries and functional
  modules.
\newblock {\em Physical review letters}, 110(17):174102, 2013.

\bibitem{golubitsky2003symmetry}
Martin Golubitsky and Ian Stewart.
\newblock {\em The symmetry perspective: from equilibrium to chaos in phase
  space and physical space}, volume 200.
\newblock Springer Science \& Business Media, 2003.

\bibitem{golubitsky2012singularities}
Martin Golubitsky, Ian Stewart, et~al.
\newblock {\em Singularities and groups in bifurcation theory}, volume~2.
\newblock Springer Science \& Business Media, 2012.

\bibitem{NCs}
Francesco Sorrentino, Louis~M Pecora, Aaron~M Hagerstrom, Thomas~E Murphy, and
  Rajarshi Roy.
\newblock Complete characterization of stability of cluster synchronization in
  complex dynamical networks.
\newblock {\em Science Advances 2}, 2015.

\bibitem{macarthur2009spectral}
Ben~D MacArthur and Rub{\'e}n~J S{\'a}nchez-Garc{\'\i}a.
\newblock Spectral characteristics of network redundancy.
\newblock {\em Physical Review E}, 80(2):026117, 2009.

\bibitem{KI1}
O.~Kinouchi and M.~Copelli.
\newblock Optimal dynamical range of excitable networks at criticality.
\newblock {\em Nature Physics}, 2:348--352, 2006.

\bibitem{KI2}
D.~B. Larremore, W.~L. Shew, and J.~G. Restrepo.
\newblock Predicting criticality and dynamic range in complex networks: effects
  of topology.
\newblock {\em Phys. Rev. Lett.}, 106:058101, 2011.

\bibitem{KI3}
Daniel~B Larremore, Woodrow~L Shew, Edward Ott, and Juan~G Restrepo.
\newblock Effects of network topology, transmission delays, and refractoriness
  on the response of coupled excitable systems to a stochastic stimulus.
\newblock {\em Chaos: An Interdisciplinary Journal of Nonlinear Science},
  21(2):025117, 2011.

\bibitem{sage}
{SageMath} - a free open-source mathematics software.
\newblock \url{http://www.sagemath.org/index.html}.

\bibitem{knight2011internet}
Simon Knight, Huan~X Nguyen, Nick Falkner, Richard Bowden, and Matthew Roughan.
\newblock The internet topology zoo.
\newblock {\em Selected Areas in Communications, IEEE Journal on},
  29(9):1765--1775, 2011.

\bibitem{wu2007cooperation}
Zhi-Xi Wu and Ying-Hai Wang.
\newblock Cooperation enhanced by the difference between interaction and
  learning neighborhoods for evolutionary spatial prisoner's dilemma games.
\newblock {\em Physical Review E}, 75(4):041114, 2007.

\bibitem{wang2010evolutionary}
Jing Wang, Bin Wu, Xiaojie Chen, and Long Wang.
\newblock Evolutionary dynamics of public goods games with diverse
  contributions in finite populations.
\newblock {\em Physical Review E}, 81(5):056103, 2010.

\bibitem{tan2014structure}
Shaolin Tan, Jinhu Lu, Guanrong Chen, and David~J Hill.
\newblock When structure meets function in evolutionary dynamics on complex
  networks.
\newblock {\em Circuits and Systems Magazine, IEEE}, 14(4):36--50, 2014.

\bibitem{macarthur2008symmetry}
Ben~D MacArthur, Rub{\'e}n~J S{\'a}nchez-Garc{\'\i}a, and James~W Anderson.
\newblock Symmetry in complex networks.
\newblock {\em Discrete Applied Mathematics}, 156(18):3525--3531, 2008.

\end{thebibliography}

\end{document}